\documentclass[letter,twocolumn]{jpsj3}
\usepackage{graphicx}           
\usepackage{txfonts}
\usepackage{color}
\usepackage{braket}
\usepackage{xspace}
\usepackage{bm}

\newcommand{\be}{\begin{equation}}
\newcommand{\ee}{\end{equation}}
\newcommand{\bea}{\begin{eqnarray}}
\newcommand{\eea}{\end{eqnarray}}
\newcommand{\beaa}{\begin{eqnarray*}}
\newcommand{\eeaa}{\end{eqnarray*}}
\newcommand{\alphaI}{{$\alpha$-(BEDT-TTF)$_2$I$_3$}\xspace}

\title{Quantum Phase Transition in Organic Massless Dirac Fermion System $\alpha$-(BEDT-TTF)$_2$I$_3$ under pressure}

\author{Yoshinari Unozawa$^1$, Yoshitaka Kawasugi$^1$, Masayuki Suda$^2$, Hiroshi M. Yamamoto$^3$, Reizo Kato$^4$, Yutaka Nishio$^1$, Koji Kajita$^1$, Takao Morinari$^5$\thanks{morinari.takao.5s@kyoto-u.ac.jp}, and Naoya Tajima$^1$\thanks{naoya.tajima@sci.toho-u.ac.jp}}

\inst{
$^1$Department of Physics, Toho University, Funabashi, Chiba 274-8510, Japan\\ 
$^2$Department of Molecular Engineering Graduate School of Engineering, Kyoto University, Kyoto 615-8510, Japan\\
$^3$Institute for Molecular Science, Okazaki, Aichi 444-8585, Japan \\
$^4$RIKEN, Hirosawa 2-1, Wako-shi, Saitama 351-0198, Japan\\
$^5$Course of Studies on Materials Science, Graduate School of Human and Environmental Studies, Kyoto University, Kyoto 606-8501, Japan 
}

\abst{
  We investigate the effect of strong electronic correlation
  on the massless Dirac fermion system, $\alpha$-(BEDT-TTF)$_2$I$_3$,
  under pressure.
  In this organic salt, one can control the electronic correlation
  by changing pressure and access the quantum critical point
  between the massless Dirac fermion phase and the charge ordering phase.
  We theoretically study the electronic structure of this system
  by applying the slave-rotor theory and 
  find that the Fermi velocity decreases without creating a mass gap
  upon approaching the quantum critical point from the massless Dirac fermion phase.
  We show that the pressure-dependence of the Fermi velocity
  is in good quantitative agreement with the results of the experiment
  where the Fermi velocity is determined by the analysis of
  the Shubnikov-de Haas oscillations in the doped samples.
  Our result implies that the massless Dirac fermion system exhibits
  a quantum phase transition without creating a mass gap
  even in the presence of strong electronic correlations.
}

\allowdisplaybreaks

\usepackage{color}

\usepackage{ulem}

\begin{document}
\maketitle
The discovery of graphene\cite{Novoselov2004} has opened up
a new era of research of massless Dirac fermions
in condensed matter physics.\cite{CastroNeto2009}
In a system with the massless Dirac fermion spectrum,
the conduction and valence bands touch at discrete points
called Dirac points, in the Brillouin zone,
and the low-energy excitations are described
by a relativistic Dirac equation,
where the velocity of light is replaced by the Fermi velocity.
When the Dirac points exist near the Fermi energy,
the conduction electrons have extremely high mobility due to the Berry phase effect.\cite{Ando1998}
Reflecting the characteristic Landau-level structure associated with
Dirac fermions, a half-integer quantum Hall effect is observed
under magnetic field.\cite{Novoselov2005,Zhang2005} 
Experimentally, massless Dirac fermions are realized
in various systems,\cite{Wehling2014}
including the surfaces of three-dimensional topological insulators,\cite{Hasan2010}
and even in the bulk of three-dimensional systems
.\cite{Armitage2018}

After experimental observation of the intriguing physical properties of massless Dirac fermions,
the next important question to address is the effect of electron-electron
interactions on massless Dirac fermions.
The interaction includes both 
long-range and short-range parts.
%
%
In contrast to conventional metals,
systems hosting Dirac fermions experience long-range Coulomb interaction 
when the Fermi energy is close to the Dirac point
because the screening effect is negligible.
The renormalization group theory shows that
the Fermi velocity is renormalized and logarithmically divergent
due to the long-range Coulomb interaction,
though the system flows to a non-interacting theory.\cite{Kotov2012}
Experimentally, a large enhancement in the Fermi velocity has been
confirmed in graphene\cite{Elias2011} by approaching
the Dirac point.
Although the long-range part of the Coulomb interaction leads to
reshaping of the Dirac cone, it does not lead to new phases.\cite{Elias2011}

%

The most enigmatic is the short-range Coulomb interaction.
The analysis of excitonic and Cooper pairing instabilities suggests
that a mass gap is created in the Dirac fermion spectrum
when the Coulomb interaction exceeds a critical value.\cite{Khveshchenko2009}
Meanwhile, theoretical analysis of
the Hubbard model on a honeycomb lattice with long-range Coulomb interactions
suggests that there is a quantum critical point
while going from a Dirac semimetal to an antiferromagnetic Mott insulator,
and that there is a decrease in the Fermi velocity due to
the on-site Coulomb repulsion.\cite{Tang2018}
While suspended graphene appears to be in the weak coupling regime,
the interaction can be enhanced in graphene on metallic substrates.\cite{Tang2018}
However, it is still challenging to drive the system
into the strong coupling regime for graphene.

In order to investigate the effect of short-range interaction on massless Dirac fermions,
the organic charge-transfer salt
$\alpha$-(BEDT-TTF)$_2$I$_3$ under pressure forms
an ideal platform.\cite{Kajita2014}
(Here, BEDT-TTF is bis(ethylenedithio)tetrathiafulvalene.)
A single crystal of $\alpha$-(BEDT-TTF)$_2$I$_3$ consists of conductive layers
of BEDT-TTF molecules and insulating layers of I$_3^-$ anions\cite{Bender1984}.
There are four BEDT-TTF molecules per unit cell.
According to the band calculation of this material under high pressure
by Katayama et al.\cite{Katayama2006},
the electronic structure is described by massless Dirac fermions.
The first band (highest-energy band; valence band)
and the second band (conduction band)
touch each other
at two points in the first Brillouin zone
where the bands exhibit linear energy dispersion.
There are two more bands, but they are energetically well separated 
from the linearly dispersing bands.
Since there is one hole per two molecules,
the HOMO-band is 3/4-filled. 
As a result, the Fermi energy is located exactly at the Dirac point.

The salient feature of this system is that one can control the electronic correlation
by changing pressure.
Under ambient pressure, this system exhibits a metal-insulator transition at 135 K,
where the insulating state is a charge-ordered insulator with a horizontal
charge stripe.\cite{Kino1995, Seo2000, Takano2001}
%
The strong electronic correlations that are responsible for forming the charge stripe
are suppressed under high pressure, and the system becomes
a massless Dirac fermion system above 1.5 GPa.\cite{Kajita2014,Tajima2000,Tajima2006,Tajima2007,Liu2016,Uykur2019}
%
The interaction effect on the massless Dirac fermions in this system has been
investigated theoretically and experimentally .\cite{Hirata2016,Hirata2017}
Like in graphene, Dirac cone reshaping has been observed in $\alpha$-(BEDT-TTF)$_2$I$_3$
using site-selective nuclear magnetic resonance.\cite{Hirata2016}
Because of the presence of tilt of the Dirac cone,
the reshaping was found to be anisotropic.
The effect of the short-range Coulomb repulsion was also
observed as the ferrimagnetic spin polarization.
Besides, a significant violation from the Korringa law suggests that the system is
in the strong coupling regime.\cite{Hirata2017}

In this study, we investigate
the quantum phase transition
in the massless Dirac fermion system, \alphaI.
We focus on the quantum phase transition
occurring upon varying the strength of the electronic correlation
by controlling the pressure in this system.
Applying the slave-rotor formalism,\cite{Florens2004}
we theoretically show that the Fermi velocity of the massless Dirac fermions
decreases upon approaching the quantum critical point
without creating a mass gap.
This theoretical result is in good quantitative agreement
with the those obtained from the experiment where the Fermi velocity is determined by the analysis
of the Shubnikov-de Haas (SdH) oscillations.

\begin{figure}[htbp]
  \begin{centering}
    \includegraphics[width=0.33\textwidth, angle=270,clip]{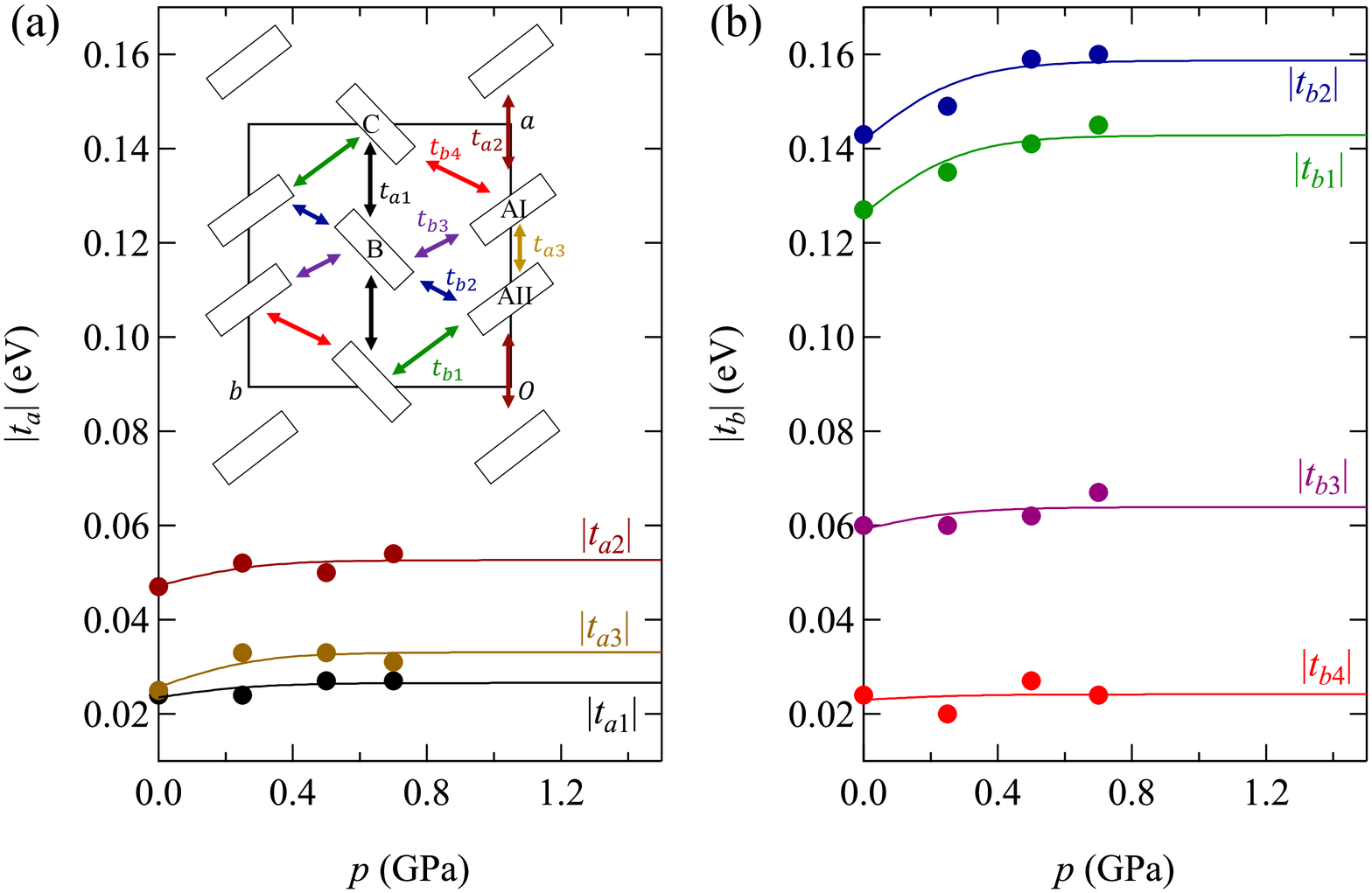}
    \caption{
      \label{fig:tatb}
      (Color online) 
      Pressure dependence of the transfer energies
      along the stacking direction of the molecules (a)
      and other directions (b).
      (inset)
      Configuration and the transfers between BEDT-TTF molecules
      in the BEDT-TTF plane in \alphaI.
      The unit cell is denoted by a square.
      There are four types of
      BEDT-TTF molecules, labeled A, A', B, and C.
      The horizontal line is taken as the $b$-axis
      and the vertical line is taken as the $a$-axis.
    }
  \end{centering}    
\end{figure}

We consider an extended Hubbard model
describing \alphaI.
The Hamiltonian is given by\cite{Katayama2006,Kobayashi2007,Kajita2014}
\bea
H &=&
\sum\limits_{\alpha ,\beta } {\sum\limits_{i,j}
  {\sum\limits_\sigma  {{t_{\alpha i,\beta j}}c_{\alpha i\sigma }^\dagger
      {c_{\beta j\sigma }}} } } \nonumber \\
& & +
\sum\limits_{\alpha ,\beta } {\sum\limits_{i,j}
  {\sum\limits_{\sigma ,\sigma '} {{V_{\alpha i,\beta j}}
      c_{\alpha i \sigma}^\dagger c_{\beta j \sigma'}^\dagger
      {c_{\beta j \sigma'}}{c_{\alpha i \sigma}}} } }
\eea
Here, $\alpha$ and $\beta$ denote four BEDT-TTF molecules,
A (AI), A' (AII), B, and C
in the unit cells as shown in the inset of Fig.~\ref{fig:tatb}(a).
The unit cells are indexed by $i$ and $j$.
$t_{\alpha i,\beta j}$ denote
the $\pi$-electron transfer energies between $\alpha$ molecule in the $i$-th unit cell
and $\beta$ molecule in the $j$-th unit cell.
$c_{\alpha i\sigma }^\dagger$ is the creation operator of the electron
with spin $\sigma = \uparrow, \downarrow$
at $\alpha$ molecule in the $i$-th unit cell.
The interaction between the electrons is described by the second term.
$V_{\alpha i,\alpha i} \equiv U/2$ is the on-site Coulomb repulsion.
$V_{\alpha i,\beta j}$ with $\alpha \neq \beta$ and $i \neq j$
describe the nearest neighbor Coulomb repulsion.
%

After the Fourier transform and introducing
the charge mean fields,
we obtain
\be
   {n_{\alpha \sigma }} = \frac{1}{N_u}\sum\limits_j
   {\left\langle {{n_{j\alpha \sigma }}} \right\rangle }
   = \frac{1}{N_u}\sum\limits_{\bf{k}} {\left\langle {{n_{{\bf{k}}\alpha \sigma }}}
     \right\rangle },
   \ee
with $N_u$ being the number of unit cells and $\bm{k}=(k_x,k_y)$
being the two-dimensional wave vector,
we obtain the following Hamiltonian:
\be
H = \sum\limits_{{\bf{k}},\sigma } {c_{{\bf{k}}\sigma }^\dagger
  {H_{\bf{k}}}{c_{{\bf{k}}\sigma }}}.
\ee
Here, 
$c_{{\bf{k}}\sigma }^\dagger
= \left( {c_{{\bf{k}}A\sigma }^\dagger ,c_{{\bf{k}}A'\sigma }^\dagger,
  c_{{\bf{k}}B\sigma }^\dagger ,c_{{\bf{k}}C\sigma }^\dagger } \right)$.
The explicit forms of the matrix elements of $H_{\bm{k}}$ are,
$(H_{\bm{k}})_{12}=
{{t_{a3}} + {t_{a2}}{e^{i{k_y}}}}$,
$(H_{\bm{k}})_{13}=
{{t_{b3}} + {t_{b2}}{e^{i{k_x}}}}$,
$(H_{\bm{k}})_{14}=
{{t_{b4}}{e^{i{k_y}}} + {t_{b1}}{e^{i{k_x} + i{k_y}}}}$,
$(H_{\bm{k}})_{23}=
{{t_{b2}} + {t_{b3}}{e^{i{k_x}}}}$,
$(H_{\bm{k}})_{24}=
{{t_{b1}} + {t_{b4}}{e^{i{k_x}}}}$,
$(H_{\bm{k}})_{34}=
{{t_{a1}} + {t_{a1}}{e^{i{k_y}}}}$.
The indices for the transfer energies are defined
in the inset of Fig.~\ref{fig:tatb}.
The diagonal matrix elements are
$(H_{\bm{k}})_{\alpha\alpha}=
U n_{\alpha\overline{\sigma}}
+ 2 \sum_{\beta(\neq \alpha)} V_{\alpha \beta}
(n_{\beta\uparrow} + n_{\beta\downarrow})$,
where $\overline{\sigma}$ denotes flipped $\sigma$
and $V_{\alpha \beta} \neq 0$ for the two nearest neighbor molecules.
When the two molecules are aligned parallel to the $y$-axis,
$V_{\alpha \beta} = V_c$,
and $V_{\alpha \beta} = V_p$ for the other directions.

The transfer energies are pressure-dependent.
Their values were evaluated from the extended H\"{u}ckel method
using the atomic coordinates obtained by crystal structure analyses
of \alphaI under hydrostatic pressure. \cite{Kondo2009}
On the other hand, the pressure dependence of the lattice constants
appear to be saturated at high pressures.\cite{Tamura2002}
Thus, we consider the following formula for the pressure dependence
of the transfer energies:
\be
   {t_\ell } = {A_\ell }
   \left[ {1 + {B_\ell }\tanh \left( {C p} \right)} \right].
\ee
Here, $\ell=\rm{a1}, \rm{a2}, ..., \rm{b4}$;
and $A_\ell$, and $B_\ell$ are fitting parameters.
The parameter $C$ is associated with the saturation of the
transfer energies.
The result of fitting is shown in Fig.~\ref{fig:tatb}
with taking $1/C=0.3$ GPa.

Now we apply the slave-rotor theory\cite{Florens2004}
to the system to investigate the effect of strong electronic correlations.
The electron creation operator at molecule $\alpha$ in the $j$-th unit cell
and spin $\sigma$ can be rewritten as
\be
c_{\alpha j\sigma }^\dagger  = {e^{i{\theta _j}}}f_{\alpha j\sigma }^\dagger
   \ee
   Note that the phase field $\theta_j$ is independent of $\alpha$.
   When electrons are created by this operator in the vacuum,
   the total number of electrons is obtained by
   applying the operator
   ${L_j} =  - i\partial /\partial {\theta _j}$.
   The advantage of this formalism is that one can distinguish
   between the Mott insulator phase and the metallic phase.
   In terms of the electron number operator,
   we cannot distinguish between them.
   However, the expectation value of $\theta_j$ vanishes
   for the Mott insulating phase while it is nonzero
   for the metallic phase.

In the salve-rotor formalism,
each electron hopping term can be rewritten as
$t_{\alpha i,\beta j}{e^{i\left( {{\theta _i} - {\theta _j}} \right)}}
f_{\alpha i\sigma }^\dag {f_{\beta j\sigma }}$.
We introduce an approximation to decouple
the fermion fields and the phase fields as
\be
\left\langle {{e^{i\left( {{\theta _i} - {\theta _j}} \right)}}} \right\rangle f_{\alpha i\sigma }^\dag {f_{\beta j\sigma }} + \left\langle {f_{\alpha i\sigma }^\dag {f_{\beta j\sigma }}} \right\rangle {e^{i\left( {{\theta _i} - {\theta _j}} \right)}}.
\ee
In the phase field system, which is now described by
a quantum XY-model, we introduce another mean field approximation,\cite{Florens2004}
$\langle {{e^{i\left( {{\theta _i} - {\theta _j}} \right)}}} \rangle
\simeq {\langle {\cos {\theta _i}} \rangle}
       {\langle {\cos {\theta _j}} \rangle}$.
As a result, the off-diagonal terms of the mean field Hamiltonian are rescaled as
$(H_{\bm{k}})_{ij} \rightarrow Z (H_{\bm{k}})_{ij}$ ($i\neq j$).
Here, $Z$ describes the band renormalization and is given by $Z=\eta^2$
with $\eta=\langle \cos \theta \rangle_{\theta}$.
The final mean field Hamiltonian is $H_f + H_{\theta}$,
where
\be
   {H_f} = \sum\limits_{{\bf{k}},\alpha ,\beta ,\sigma }
   {c_{{\bf{k}}\alpha \sigma }^\dag \left[ {Z\left( {{H_{\bf{k}}}} \right)_{\alpha\beta} 
         \left( {1 - {\delta _{\alpha \beta }}} \right)
         + \left( {H_{\bf{k}}} \right)_{\alpha\alpha}
         {\delta _{\alpha \beta }}} \right]
     {c_{{\bf{k}}\beta \sigma }}}.
   \ee
The expectation value $\langle \cos \theta \rangle_{\theta}$
is computed based on the following Hamiltonian:
\be
   {H_\theta } = \frac{U_{\rm eff}}{2}{{\widehat{L}}^2} + 2\chi \eta \cos \theta,
\ee
where
${\widehat{L}} =  - i\partial/\partial \theta$.
Here,
$\chi  = \sum\limits_{\alpha ,{\bf{k}},\sigma }
{\left\langle {f_{\alpha {\bf{k}}\sigma }^\dag {f_{\alpha {\bf{k}}\sigma }}}
  \right\rangle } /N_u$.
$U_{\rm eff} \sim U$ is the effective short-range interaction.
The expectation value
$\left\langle {f_{\alpha {\bf{k}}\sigma }^\dag {f_{\alpha {\bf{k}}\sigma }}}
  \right\rangle$
  is computed based on the Hamiltonian, $H_f$.
We note that
there is a Lagrange multiplier enforcing
the constraint on the expectation value of
$\langle {\widehat{L}}\rangle_{\theta}$.
However, this parameter can be removed
by shifting the origin of the phase field.
The values $\eta$ and $\chi$
are to be determined self-consistently.

\begin{figure}
\begin{centering}
  \includegraphics[trim=10 53 5 55, width=0.38\textwidth, angle=270,clip]{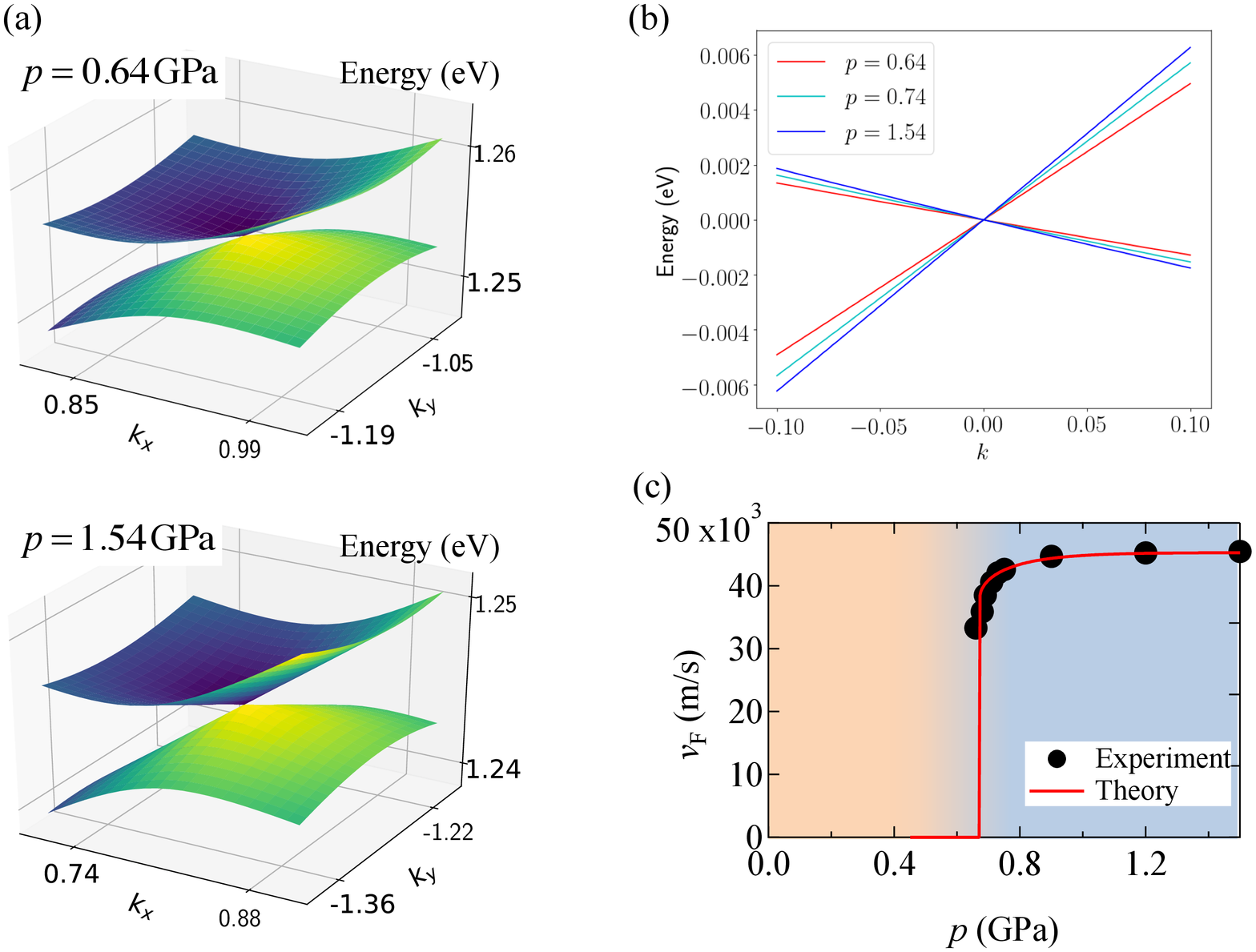}
\caption{\label{fig2}(Color online)
  (a) Band structure in the vicinity of the Dirac node
  for $p = 0.64, 1.54$~GPa.  
  (b) Energy dispersion around the Dirac node in the direction of
  the $k_x$ axis
  for $p = 0.64, 0.74, 1.54$~GPa.
  The Fermi velocity decreases as we decrease $p$.
  (c) Pressure dependence of the Fermi velocity for the experiment (closed circles) and theory (solid line).
  The interaction parameters are $U=0.4$, $V_c=0.17$, $V_p=0.05$,
  and $U_{\rm eff}=0.35$ in units of eV.
}
\end{centering}
\end{figure}

In Fig.~\ref{fig2}(a), we show the band structure
around the Dirac nodes at $\tilde{p} \equiv p-p_0 = 0.3, 1.2$~GPa.
Here, $p_0 = 0.336$~GPa denotes the shift of the pressure
due to the suppression of the transfer energies associated with
the strong electronic correlation effect.
The Fermi velocity $v_{\rm F}$
decreases as we decrease $\tilde{p}$
as seen in Fig.~\ref{fig2}(b).
This decrease in $v_{\rm F}$ is associated with a decrease in $Z$
that reflects the effect of the strong electronic correlation,
and we may adopt a simple formula $v_{\rm F} = Z v_0$ for $v_F$
with $v_0$ being a parameter.
We note that the decrease in $v_{\rm F}$ due to the strong electronic correlation
is consistent with the previous studies.\cite{Tang2018,Hirata2016}
The pressure dependence of $v_{\rm F}$ is compared
with the experiment in Fig.~\ref{fig2}(c).
The theory is in good quantitative agreement with the experiment.
In the experiment, $v_{\rm F}$ is determined by analyzing the SdH oscillations as described below.
Since the tilt of the Dirac cones in this system,
$v_{\rm F}$ is given by the average Fermi velocity
$v_{\rm F} (\phi)$, defined in the direction with an azimuthal angle $\phi$,
as $\int_{0}^{2\pi} v_{\rm F}(\phi)^{-2} d\phi = 2\pi v_{\rm F}^{-2}$.

Now we describe the experiment.
To investigate the electronic structure of the system,
we observed SdH oscillations under pressure
between 0.66 GPa and 1.5 GPa at 0.5 K.
To detect the SdH oscillations, we need to shift $E_{\rm F}$ from the Dirac point (DP).
In this experiment, we injected holes into $\alpha$-(BEDT-TTF)$_2$I$_3$
by the contact electrification method\cite{Tajima2013,Tajima2018a} as follows.
The carrier density per layer of $\alpha$-(BEDT-TTF)$_2$I$_3$
under high pressure and at low temperatures is estimated to be approximately
10$^8$ cm$^{-2}$.
Thus, the effects of hole doping can be detected by transport measurement
by fixing a thin crystal onto a polyethylene naphthalate (PEN) substrate,
which is weakly negatively charged.
The thickness of the crystal measured with a step profiler was approximately 100 nm.
Number of layers (pairs of BEDT-TTF molecular layers and I$_3^-$
anion layers; vide infra)
is estimated to be approximately 57.
Note that holes are injected into a few layers
and rapidly decrease from the substrate.
Thus, the observed SdH oscillations are associated with
the first layer on the substrate in this experiment.

\begin{figure}
\begin{centering}
  \includegraphics[trim=10 20 20 20, width=5.5cm, angle=270,clip]{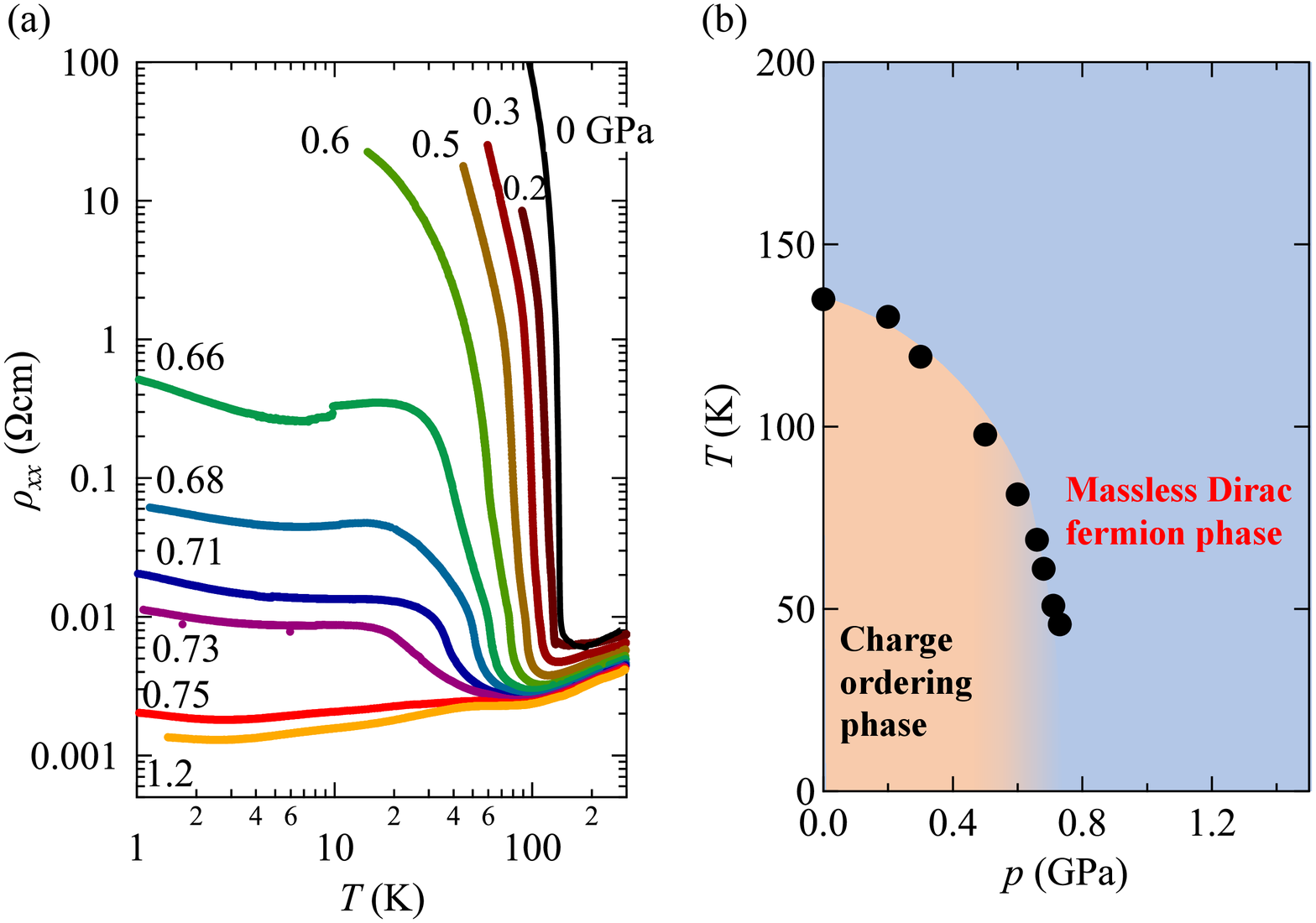}
\caption{\label{fig3}(Color online) 
  (a) Temperature dependence of $\rho_{xx}$
  in $\alpha$-(BEDT-TTF)$_2$I$_3$ fixed on the PEN substrate under pressure.
  (b) Temperature-pressure phase diagram.
  }
\end{centering}
\end{figure}

The critical pressure for the quantum phase transition
was determined from the resistivity, $\rho_{xx}$, measurement.
We show the temperature dependence of $\rho_{xx}$
for different pressures in Fig.~\ref{fig3}(a).
The system evolves from the insulating state into
a metallic state as we increase the pressure.
From this measurement,
the critical pressure is estimated to be approximately 0.75 GPa.
At intermediate pressures, 
the system consists of the insulating phase
and the massless Dirac electron phase
due to the inhomogeneity of the pressure effect.
The temperature-pressure phase diagram is shown in Fig.~\ref{fig3}(b)
for $\alpha$-(BEDT-TTF)$_2$I$_3$ on the PEN substrate.
This phase diagram should be compared with that of the bulk thick crystals of the same system.
The softness of the PEN substrate results in a large pressure effect
on the thin crystal such that the massless Dirac electron system
is realized in about half of the necessary pressure
needed in the case of a bulk thick crystal.

\begin{figure}
\begin{centering}
  \includegraphics[trim=50 20 50 20, width=5.3 cm, angle=270, clip]{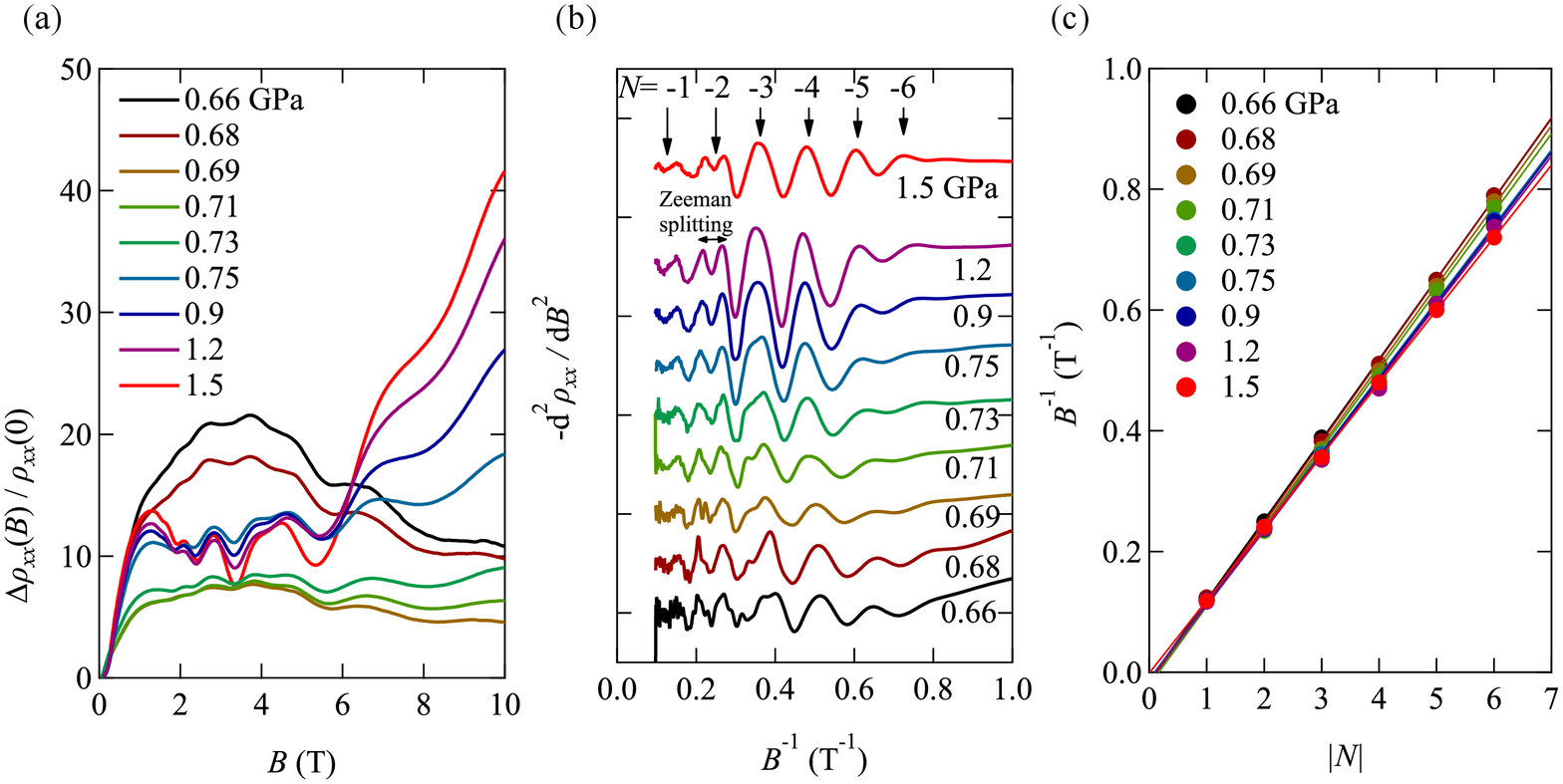}
\caption{\label{fig4}(Color online)
  (a) Magnetic field dependence of $\rho_{xx}$ under several pressures at 0.5 K.
  (b) Second-order differential of $\rho_{xx}$, $-(d^2\rho_{xx}/dB^2)$,
  as a function of $B^{-1}$. The periodic oscillations indicate the SdH signal.
  The oscillation maxima are denoted by the Landau index $N$.
  The Zeeman effect splits $N$=-2 and -1 Landau levels.
  (c) Landau index dependence of the value of $B^{-1}$
  for the SdH oscillation maxima.
  In the case of Dirac electron systems, the linear extrapolation
  of the data to $B^{-1}=0$ should be 0. 
 }
\end{centering}
\end{figure}

We confirmed that the system is in the Dirac electron phase
from the Berry phase analysis,
where the circular orbits around the DP result in the Berry phase $\phi_B=\pi$.
The magnetic field, $B$, dependence of $\rho_{xx}$
for different pressures is shown in Fig.~\ref{fig4}(a).
The oscillation pattern becomes clearer by taking
the second-order differentials of $\rho_{xx}$
as a function of $B^{-1}$
as shown in Fig.~\ref{fig4}(b).
We apply the semi-classical magneto-oscillation analysis
to find the phase of the SdH oscillation.
We note that the maxima of the oscillations
are associated with the Landau levels
denoted by $N$ in Fig.~\ref{fig4}(b). 
Because of the Zeeman energy,
the Landau levels with $N=-1$ and $N=-2$ are lifted.
Using this splitting,
we show the linear approximation formula
$B^{-1}=\frac{2\pi^2}{\phi_0 S_F}(N+\gamma)$ in Fig.~\ref{fig4}(c).
Here, $S_{\rm F}=2\pi^2B_f\phi_0^{-1}$ is the Fermi surface cross-section area,
$B_f$ is the SdH frequency, and
$\phi_0=h/2e=2.0678\times10^{-15}$ Wb is the fluxoid.
The Onsager phase factor is defined by $\gamma \equiv 1/2-\phi_B/2\pi$.
For the case of conventional metals,
the Landau level energies are given by
$E_N=\hbar \omega_c (N+1/2)$ with $\omega_c$ as the cyclotron frequency
and we obtain $\gamma = 1/2$.
Meanwhile, we have $E_N=\pm \sqrt{2e\hbar v_{\rm F}^2 |N||B|}$ 
for the massless Dirac electron systems
and we obtain $\gamma=0$.
Figure~\ref{fig4}(c) shows that $\gamma \simeq 0$.
Therefore, the system is in the Dirac electron phase with $\phi_B=\pi$
under all applied pressures.

Experimentally, the Fermi velocity, $v_{\rm F}$,
is determined from the splitting of the Landau levels
due to the Zeeman energy, that is,
$E_{N\uparrow \downarrow}(B)=E_N\pm g\mu_B B/2$ with $g=2$.
For instance, $v_{\rm F}$ is obtained from the relation 
$E_{-2\uparrow}(B_{-2\uparrow})=E_{-2\downarrow}(B_{-2\downarrow})$.
Here, $B_{-2\uparrow}$ and $B_{-2\downarrow}$ are magnetic field strengths, 
where the up-spin and down-spin Landau levels for $N=-2$ cross the Fermi energy. 
Pressure dependence of $v_{\rm F}$ is shown in Fig.~\ref{fig2}(b). 
The system approaches the quantum critical point
with a rapid decrease in $v_{\rm F}$ without creating a mass gap.

To conclude, we have investigated the effect of strong electronic correlations 
on the massless Dirac fermion system $\alpha$-(BEDT-TTF)$_2$I$_3$ under pressure.
We have theoretically shown that the Fermi velocity $v_{\rm F}$ decreases
as we increase the electronic correlation
without creating the mass gap.
The change in $v_{\rm F}$ was obtained by applying the slave-rotor theory.
The result is in good quantitative agreement
with the experimentally obtained $v_{\rm F}$,
where $v_{\rm F}$ is determined from the analysis of the SdH oscillations
of doped samples.
Our study suggests that there is no mass gap opening
at the quantum critical point between the massless Dirac fermion phase
and charge ordering phase.
To the best of our knowledge, there are no other massless Dirac fermion systems
where one can reach the strong electron correlation regime.
Thus, it will be interesting to further explore the physics around this quantum critical point
in $\alpha$-(BEDT-TTF)$_2$I$_3$.

\acknowledgments{
  This work was supported by MEXT/JSPJ KAKENHI under
  Grant Nos. 16H06346 and 20K03870.
  The authors are grateful to the Equipment Development Center,
  the Institute for Molecular Science, for the technical assistance
  provided by the Nanotechnology Platform Program (Molecule and
  Material Synthesis) of the Ministry of Education Culture, Sport,
  Science and Technology (MEXT), Japan.
}

\bibliography{../../../refs/lib}

\end{document}